# In situ monitoring of block copolymer self-assembly via solvent exchange through controlled dialysis with light and neutron scattering detection

Martin Fauquignon,* Lionel Porcar,* Annie Brûlet, Jean-François Le Meins, Olivier Sandre, Jean-Paul Chapel, Marc Schmutz, and Christophe Schatz*

**ABSTRACT:** Solution self-assembly of amphiphilic block copolymers (BCs) is typically performed by solvent to water exchange. However, BC assemblies are often trapped in metastable states depending on the mixing conditions, such as the magnitude and rate of water addition. BC self-assembly can be performed under near thermodynamic control by dialysis which accounts for a slow and gradual water addition. In this letter we report the use of a specifically designed dialysis cell to continuously monitor by dynamic light scattering and small-angle neutron scattering the morphological changes of PDMS-*b*-PEG BCs self-assemblies during THF-to-water exchange. The complete phase diagrams of near-equilibrium structures can then be established. Spherical micelles first form before evolving to rod-like micelles and vesicles, decreasing the total developed interfacial area of self-assembled structures in response to increasing interfacial energy as the water content increases. The dialysis kinetics can be tailored to the time scale of BC self-assembly by modifying the membrane pore size, which is of interest to study the interplay between thermodynamics and kinetics in self-assembly pathways.

Amphiphilic block copolymers (BCs) in solution can generate a variety of self-assembled structures depending on their block length and chemical composition.[1–4] Compared to surfactants which self-assemble into micellar structures at equilibrium, BC assemblies are often trapped in long-lived metastable states due to slow polymer chain dynamics associated with large molecular weight, hydrophobicity or glassy state of polymers.[5–13] BC self-assembly can be facilitated by dissolving the block copolymer in a good solvent for both blocks before adding a selective solvent, most often water, which is good for one block and poor for the other. The resulting morphologies are related to the water content in solvent mixtures and polymer concentration, which can be depicted by composition-morphology maps obtained through turbidity or electron microscopy observations[14,15] and by computer simulation,[16–18] in particular coarse-grained methods. A recurring issue with the solvent exchange method is the equilibrium state of BC aggregates, which depends not only on the final water content but also on the magnitude and rate of the water jump.[14,19–21] In order to promote the formation of near-equilibrium morphologies, as expected to build phase diagrams, water should be added slowly and gradually, by analogy with the conditions used for thermal annealing in bulk and thin film studies.[22] A fast and large water jump can lead to kinetically arrested morphologies, which is the basis of the flash nanoprecipitation technique.[23,24] In order to achieve slow and homogeneous mixing conditions, the mixing process should preferably rely on low Peclet Number solvent inter-diffusion rather than turbulent convection.[25] Microfluidic technology based on the principle of flow focusing,[26] multilamination[27,28] or chaotic laminar mixing[28,29] allows for such mixing conditions, but is inherently limited to short self-assembly time scales (< 1-10 s), which is insufficient for some BC systems to reach their final equilibrium morphology. For example, depending on the water content in the medium, the equilibration of BC vesicles may take several minutes and even hours due to slow reorganization processes.[14,19,25,30] Compared to microfluidic techniques, dialysis, which also relies on the principle of solvent interdiffusion, remains the simplest approach to achieve near-equilibrium BC self-assembly without time-scale restrictions.[31–33] However, morphology changes of BC structures can be hardly monitored in real time using dialysis bags or tubings unless sample aliquots are collected for further analysis. Yet, this approach suffers from several limitations: i) the polymer concentration cannot be kept constant during regular dialysis, ii) morphology changes occurring over narrow ranges of solvent composition can be overlooked, iii) sampling disrupts dialysis equilibrium, iv) physical aging of collected samples before any analysis is inevitable.

In this study, we performed for the first time kinetic monitoring of BC self-assembly during slow solvent exchange using a dedicated dialysis cell designed for *in situ* time-resolved light and small-angle neutron scattering experiments (Figure 1A). The main objective of this letter is to highlight the capability and relevance of the dialysis-scattering coupling approach to reveal the intrinsic properties of BC self-assembly under near-equilibrium conditions. Only under these circumstances, good understanding of the physical phenomena can be captured.

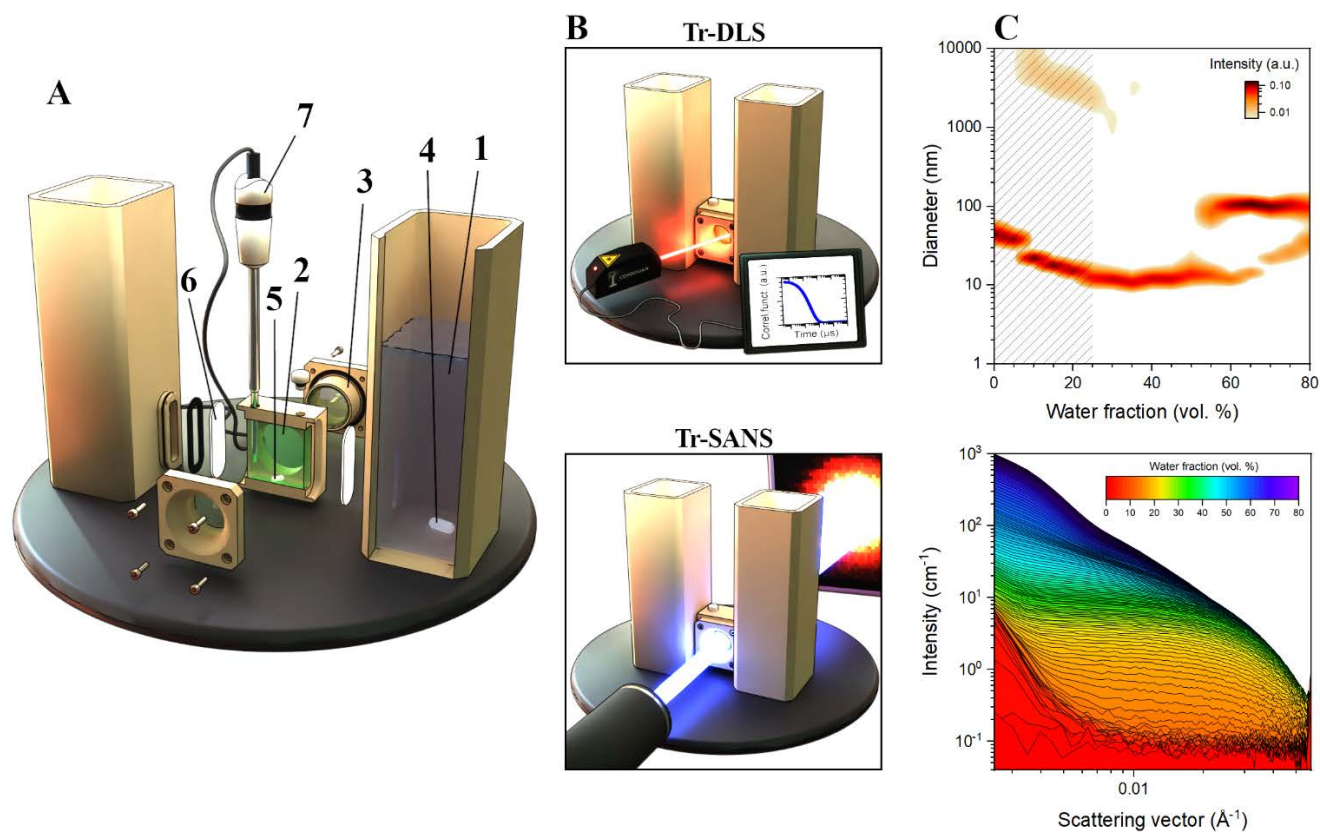

**Figure 1.** (A) Exploded view of the dialysis setup used for continuous kinetic monitoring of BC self-assembly: ①  Lateral reservoirs containing water (2 × 250 mL), ② Scattering cell containing the BC solution in THF (V = 4.5 mL), ③ Quartz windows (thickness = 1 or 2 mm), ④ Reservoir stirring, ⑤ Cell stirring, ⑥ Custom-cut ultrafiltration disc in regenerated cellulose of defined porosity (× 2). ⑦ Conductivity probe. All plastic parts are made in solvent-resistant PEEK. (B) Configuration of the remote DLS head (*top*) and SANS (*bottom*) detection. (C) Tr-DLS and Tr-SANS mapping of the dialysis of PDMS27-*b*-PEG17 at 5 g/L in THF against water: intensity-weighted size distribution (hydrodynamic diameters) obtained by a SBL inverse Laplace transform analysis (*top*) and SANS curves (*bottom*) as function of the water fraction. The upturn at very low-*q* of the scattered intensity that is visible for solvent dialysis alone was systematically subtracted from the SANS curves of polymer solution dialysis (Figure S3).

Our approach is therefore complementary to fast flow mixing with SAXS or SANS detection,[34,35] previously used to study non-equilibrium properties of BC self-assembly.[36–38]

Polydimethylsiloxane-*block*-polyethylene glycol (PDMS-*b*-PEG) diblock copolymers forming vesicles upon THF exchange with water were used as model BCs. The self-assembly properties of PDMS-*b*-PEG BCs either alone or with lipids has been reported in previous works.[39,40] The critical aggregation concentration was found to be on the order of 0.01 g/L with the BC compositions used in the present work.[41] Both the high conformational flexibility (persistence length = 5.5 Å)[42] and the rubbery state of PDMS blocks ($T_g$ = -123°C)[43] should favor vesicle formation under near-equilibrium conditions.

The BC solution in THF at a concentration of 0.5% (w/v) was introduced into the scattering discoid cell, which was separated from two water-containing reservoirs by cellulose membranes (Figure 1A and Picture S1). Deuterated solvents were used for SANS experiments. The scattering cell and reservoirs were continuously stirred at low speed (50 rpm) to allow homogeneous solvent inter-diffusion. Importantly, the cell design allowed the BC concentration to be kept constant over the entire dialysis time. 1 mM NaCl was added in the reservoirs to monitor the progress of dialysis from the ionic conductivity (Figures S1A-S1B) while providing isotonicity of the system (*see* Supporting Information). By varying the membrane porosity, the almost complete solvent exchange can be achieved in less than 10 hours (100 kDa membrane) or up to over 20 hours (10 kDa membrane) (Figure S1C). The data presented in this letter were obtained with PDMS27-*b*-PEG17 (hydrophilic mass fraction, *f* = 26%) using a 10 kDa membrane for which the THF/water exchange rate is low (< 0.1 vol. % min$^{-1}$) (Figure S1D). Does this exchange rate match the structural change kinetic of the BC? In principle yes if we refer to previous works with polystyrene-*b*-polyacrylic acid BCs of large MW, for which typical relaxation times below ~ 5 min were mainly found for 1 vol.% water jumps performed at different polymer concentrations and initial water contents in the critical regions of the phase diagram where morphological transition occurs.[44,45] Data collected with another copolymer composition and membrane porosity are provided in the supplementary material and will be discussed further.

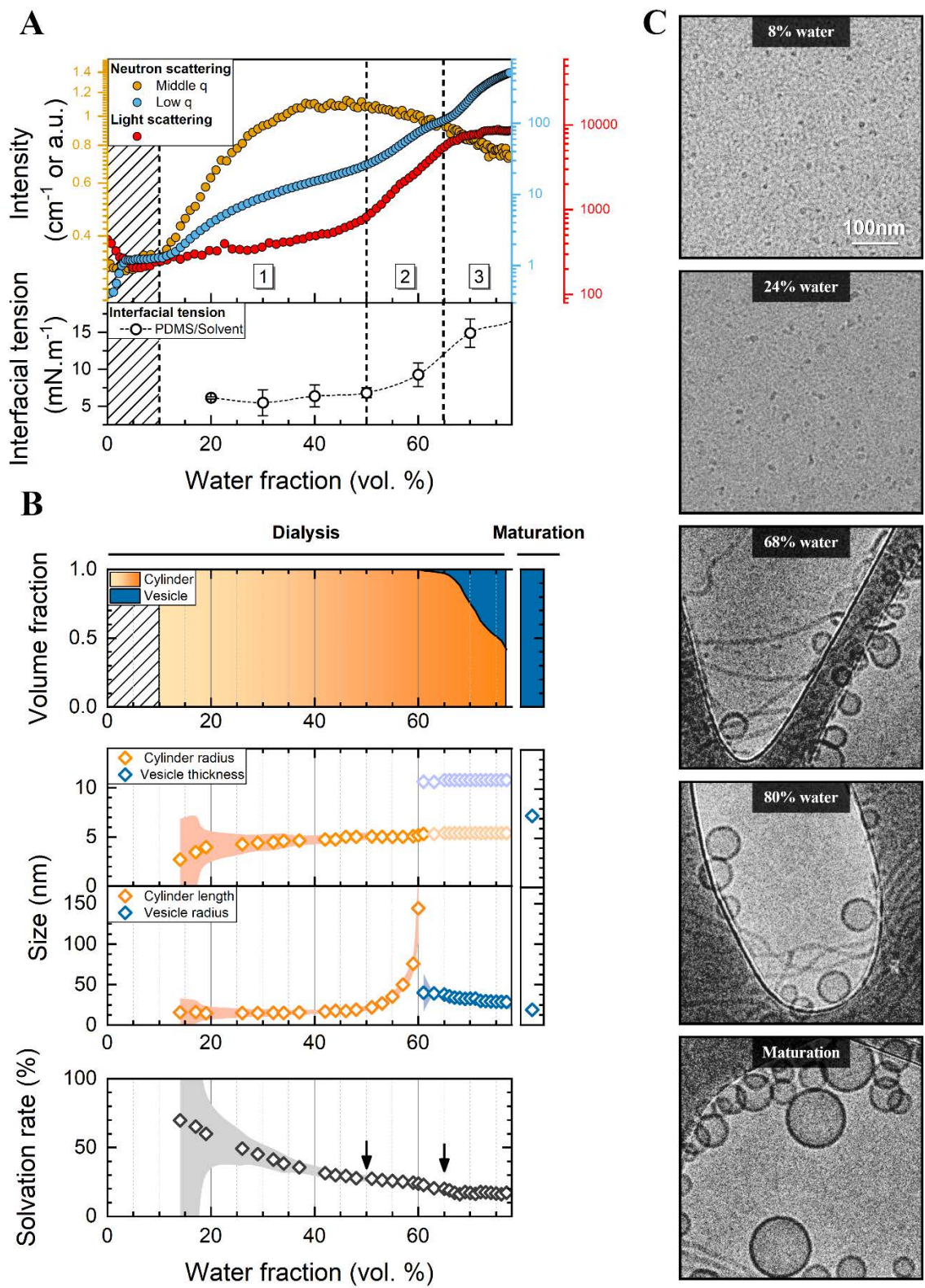

**Figure 2.** (A) Neutron scattering intensities averaged at low q (0.0028-0.0075 Å⁻¹) and middle q (0.052-0.057 Å⁻¹), light scattering intensity (q = 0.0027 Å⁻¹) and interfacial tension between PDMS and the solvent mixture (see also Figure S14) as function of water volume fraction. Three areas of interest are identified (see text). (B) Main results derived from SANS curves fitting plotted as function of the water volume fraction: i) volume fractions of the aggregate morphologies, ii) characteristic sizes of BC aggregates (for $f_w$ > 60%, light orange and light blue symbols correspond to the cylinder radius kept constant and the membrane thickness imposed as twice the radius of the rod-like micelles, in accordance with the cryo-TEM images), iii) solvation rate ($\Psi$) of PDMS blocks in BC assemblies, expressed as the volume fraction of THF-$d_8$ remaining in the core of aggregates (see Eq.18 in the supporting information). The arrows highlight small discontinuities corresponding to morphological transitions. C) Cryo-TEM analysis of the dialyzed BC solution at defined water volume fractions (see also Figures S8-S13).

A nanoparticle size analyzer combining time-resolved DLS (Tr-DLS) and *in situ* remote optical probe (VASCO KIN™, Cordouan Technologies)[46] was used for kinetic monitoring of the dialysis (Figure 1B). By providing valuable information about the hydrodynamic size and size-dispersity of BC aggregates, Tr-DLS gives a first insight into the pathway of vesicle formation (Figure 1C and Figure S2). For low water volume fractions ($f_w$ < 25%), DLS measurements were poorly resolved due to almost complete contrast matching of the PDMS block in THF ($n_{PDMS}$ = 1.407, $n_{THF}$ = 1.405). The first detectable structures were ~10 nm-diameter aggregates, most likely micellar aggregates, which developed in solution for 25% < $f_w$ < 50%. A second population of 100 nm-diameter aggregates appeared abruptly at $f_w$ = 55%. Thereafter, micelles grew and tended to fuse with the larger aggregates before the end of dialysis, which was stopped after 22 hours, corresponding to $f_w$ = 80%.

Solvent exchange was monitored under similar conditions by time-resolved small angle neutron scattering (Tr-SANS) to determine the aggregate morphology (Figure 1B). Tr-SANS measurements were performed on the D22 beamline at Institut Laue-Langevin (ILL, Grenoble).[47–49] SANS calibration of the dialysis setup was carried out by means of $H_2O/D_2O$ exchange, to check that the exchange was indeed total (unpublished results). SANS curves were acquired in 5-minute increments during dialysis (see Supporting Information), *i.e.* the water fraction increased by a maximum of 0.5 vol.% during one acquisition (Figure S1.D). SANS curves obtained during dialysis are reproduced in Figure 1C where the color shift highlights the structural changes in BC self-assembly when the water content increases (individual intensity curves are reported in Figure S6). At the very low $q$-end, the upturn of the scattered intensity for $f_w$ < 20% stems in part from long-range fluctuations in the solvent composition, as verified by dialyzing THF-$d_8$ without copolymer against $D_2O$ (Figure S3) and more importantly from the presence of some supramolecular assemblies with sizes exceeding 200 nm according to cryo-TEM analysis (Figure S8). The aggregates exhibit a cluster-like structure composed of weakly interacting BC micelles coexisting with free BC micelles, as previously reported for similar BC systems close to the critical micellar concentration.[50] Their formation at low water contents can be attributed to the low interfacial energy of the core-forming PDMS and the limited repulsive hydration forces between PEG blocks. The fact that they could not be seen by DLS was due to the almost iso-refractive conditions at the beginning of the dialysis. For $f_w$ > 20%, these supramicellar assemblies were not detected on SANS curves at low $q$ (Figure S6), indicating their metastable state. The variation in scattering intensity at low $q$ and middle $q$, after subtraction of the pure solvent contribution and ignoring the small-$q$ upturn, provides a qualitative yet informative insight into BC self-assembly. Notably, discontinuities were found in the low-$q$ SANS intensity profile and to a lesser extent in the LS intensity variation for water fractions of 10%, 50% and 65% suggesting solvent-induced morphological transitions depicted by the three regions in Figure 2A. The intensity in the mid-$q$ region increased sharply for 10 % < $f_w$ < 50%, which was attributed to BC self-assembly triggered by the PDMS desolvation (Figure 2B). Then, the decrease in intensity observed for $f_w$ > 50% might reflect a change in interface structure.

SANS curves were fitted using a combination of a small number of form factors (sphere, cylinder, vesicle) to introduce as few fitting parameters as possible (see supporting information). Only for $f_w$ < 10% the curves could not be adequately fitted due to the too low scattered intensity (Figure 2A). However, small spherical micelles ($R$ = 5 nm) could be well detected by cryo-TEM at low water contents ($f_w$ = 8%), as shown in Figure 2C. Then, short cylindrical micelles of almost constant size ($R$ = 3-5 nm, $L$ = 15 nm) were fitted in a wide range of water fractions, *i.e.* between 10% and 50%, where PDMS blocks are gradually de-solvated (Figure 2B). Cylinders appear to be rather polydisperse, both in size and shape, as seen by cryo-TEM (Figure 2c). For $f_w$ > 50% the system enters a critical zone characterized by a steep elongation of small cylinders into rod-like micelles, the length of which could be accurately determined by SANS up to 150 nm, as imposed by the configuration used in SANS (Figure 2B and Figure S5). However, cryo-TEM shows that rods reached much greater lengths, in the micron range for $f_w$ > 65%. At the same time, the first vesicles developed in solution, the proportion of which gradually increased with $f_w$ at the expense of the rods. When the dialysis was stopped after 22 hours ($f_w$ = 78%), the BC solution contained 60% vesicles and 40% rod-like micelles. The THF content in the vesicle membrane was presumably equal to that of the medium, *i.e*, 20 vol.% (Figure 2B). The solution was then stored for 40 days at room temperature. The initially bimodal size distribution (DLS) gradually turned into a monomodal distribution (Figure S7A). Only vesicles of smaller radius ($R$ = 19 nm) could be detected by SANS (Figure S7B) and cryo-TEM (Figure 2C and S13). The membrane thickness of the equilibrated vesicles ($\delta$ = 7 nm) was slightly smaller than that imposed for the fitting of the rod-vesicle mixtures during dialysis (Figure 2B).

The overall self-assembly mechanism of PDMS27-*b*-PEG17 with increasing water content is in agreement with the typical spherical micelles → rod-like micelles → vesicles transition pathway reported for several BC systems.[14,20,25] It is consistent with an increase of the interfacial tension of the core-forming PDMS as the solvent polarity increases. Micellar aggregates of spherical shape first form ($f_w$ < 10%) to minimize the total interfacial area, which in turn accounts for the core chain stretching in the radial directions due to packing constraints (localization of junctions at the interface, uniform density of PDMS segments in the core).[51–54] In the process of core enlargement by attraction of copolymers from the solution, when the entropy penalty associated with the stretching is too high, aggregates must adopt another geometry to relax the stretching and thus minimize the total free energy.[14,53] As a result, aggregates of reduced curvature (hence, higher packing density) are successively formed with increasing water content : rod-like micelles for $f_w$ > 10% and vesicles for $f_w$ > 65%. Inverted BC structures previously reported[55,56] cannot form further beyond the vesicular structure due to the reduced mobility of BC chains at very high water contents.[14]

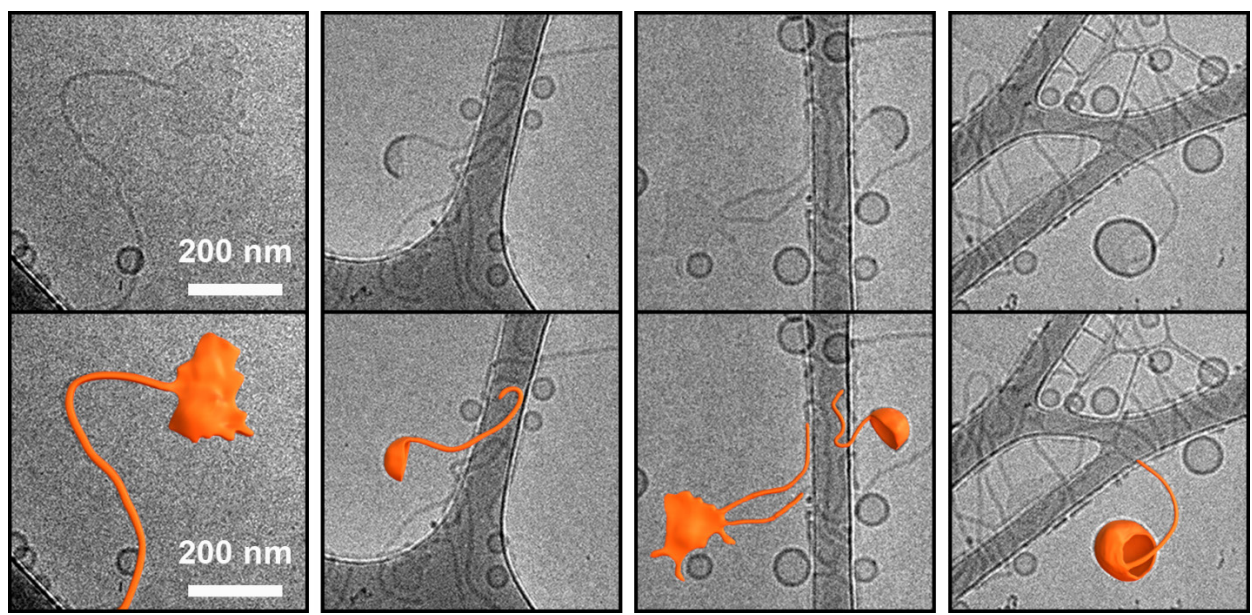

**Figure 3.** Cryo-TEM images of the PDMS27-*b*-PEG17 copolymer solution at $f_w$ = 80% illustrating the most probable transition pathway from rod-like micelles to vesicles (scale bar = 100 nm) (*top*). Highlighting of the areas of interest (*bottom*). Three different stages can be distinctly observed from left to right: i) membrane formation at the rod end-caps, ii) membrane bending, iii) membrane closure into vesicle.

The continuous monitoring of solvent exchange using LS/SANS detection and supported by cryo-TEM imaging revealed some important aspects: i) The critical water fraction at which micellization takes places ($f_w$ < 10%) is in the same range as that obtained for PS-*b*-PAA copolymers containing large PS blocks.[14] However, no jump in scattering intensity could be associated to the micellization due to the too low micelle size and isorefractive conditions in the case of LS detection. ii) The variation of SANS intensity at low *q* allows the identification of breaks at $f_w$ = 50% and $f_w$ = 65% (Figure 2A) associated to the formation of rod-like micelles and vesicles, respectively. It coincides with small discontinuities found in the variation of the PDMS solvation rate (arrows in Figure 2B). This can be further appreciated from the variation of the interfacial tension of PDMS with THF:water mixtures that shows a marked increase above $f_w$ = 50%, indicating that the rod-to-vesicle transition is indeed driven by an increase in the interfacial energy between the hydrophobic core of self-assemblies and the solvent (Figure 2B and Figure S4D). iii) The possible mechanism by which this transition occurs is shown in Figure 3 where cryo-TEM images obtained at $f_w$ = 80% depict the flattening of the rod ends into lamella, followed by the gradual bending of lamella into vesicles, as previously reported theoretically and experimentally.[20,44,57–60] iv) PDMS-*b*-PEG copolymers containing small and soft hydrophobic blocks could not fully equilibrate even after 22 hours of dialysis. The achievement of a unique vesicular morphology, as expected from the BC composition,[10] required more than one month due to slow chain dynamics and strong repulsive interaction among BC aggregates.[61] This excludes that in many situations BC self-assembly can truly reach the final equilibrium state at high water contents, especially for BCs containing large, rigid or crystalline hydrophobic blocks.[20,62,63]

The shorter PDMS23-*b*-PEG13 copolymer (*f* = 26%) followed a quite similar self-assembly pathway (Figures S15, S16, S17 and Table S3) as previously described but with marked differences regarding the vesicle volume fraction and also the vesicle dispersity, as can be seen from the well-defined oscillations in the mid-*q* region of SANS curves (Figure S15). Finally, the influence of the dialysis kinetics was investigated with PDMS27-*b*-PEG17 using a 100 kDa membrane instead (Figures S18, S19, S20 and Table S4). In such conditions, 80% of THF was exchanged by water in less than 8 hours (Figure S1C). The striking point of this faster dialysis was the formation of the first vesicles at $f_w$ = 25% compared to $f_w$ = 65% with a 10 kDa membrane. The rapid increase in interfacial energy due to fast water diffusion may exert additional constraints on the copolymer packing, inducing a shift of the critical water content required for vesicle formation.[27,63] The water exchange rate was 2 to 3 times higher than with the 10 kDa membrane at the beginning of the dialysis (< 1 h) where the polymer chain dynamics is high in relation with the large THF content (Figure S1D). Therefore, the formation of inhomogeneities cannot be completely excluded even if the cell and the reservoirs are continuously stirred. However, no kinetically trapped structure was observed, the equilibrium between rod and vesicle morphologies slowly evolving in favor of vesicles as the water fraction increased.

In conclusion, the new dialysis cell coupled with LS/SANS detection enables continuous monitoring of BC self-assembly in an evolving energy landscape, depending on the fraction and addition rate of selective solvent. The approach could be applied to other supramolecular systems controlled by the diffusion of selective solvents or any other molecular entities capable of triggering self-assembly including acids or bases, ions, H-bonding agents or small-molecule ligands.[64–69]

## ASSOCIATED CONTENT

**Supporting_Information**. This material is available free of charge via the Internet at http://pubs.acs.org.

### Corresponding Authors

**Martin Fauquignon** - *Univ. Bordeaux, CNRS, Bordeaux INP, LCPO, UMR 5629, F-33600, Pessac, France*; Email: martin.fauquignon@gmail.com

**Lionel Porcar** - *Institut Laue-Langevin (ILL), F-38042 Grenoble, France*; Email: porcar@ill.fr

**Christophe Schatz** - *Univ. Bordeaux, CNRS, Bordeaux INP, LCPO, UMR 5629, F-33600, Pessac, France*; Email: schatz@enscbp.fr

### Authors

**Annie Brûlet** - *Laboratoire Léon Brillouin, Commissariat à l'Energie Atomique et aux Energies Alternatives (CEA) Saclay, F-91191 Gif-sur-Yvette, France*

**Jean-François Le Meins** - *Univ. Bordeaux, CNRS, Bordeaux INP, LCPO, UMR 5629, F-33600, Pessac, France*

**Olivier Sandre** - *Univ. Bordeaux, CNRS, Bordeaux INP, LCPO, UMR 5629, F-33600, Pessac, France*

**Jean Paul Chapel** - *Centre de Recherche Paul Pascal (CRPP), UMR CNRS 5031, Université de Bordeaux, F-33600 Pessac, France*

**Marc Schmutz** - *Université de Strasbourg, CNRS, Institut Charles Sadron, UPR 22, F-67034 Strasbourg, France*

### Funding sources

This work was supported by the French National Research Agency (ANR), under the grant ANR 18-CE06-0016 (SISAL project).